\documentclass[pra,aps,amsmath,amssymb,notitlepage,twocolumn,floatfix,superscriptaddress]{revtex4-1}
\usepackage{amsthm}
\usepackage{amsfonts}
\usepackage{siunitx}
\usepackage{amsmath}
\usepackage{amssymb}
\usepackage{graphicx}
\usepackage{verbatim}
\usepackage[colorlinks]{hyperref}
\usepackage{tikz}
\usepackage{pgfplots}
\usepackage{braket}
\usepackage{algorithm}
\usepackage{algpseudocode}

\definecolor{linkcolor}{RGB}{0,83,166}
\hypersetup{
    colorlinks = true,
    allcolors = {linkcolor}
}

\begin{document}
\title{Emulating the coherent Ising machine with a mean-field algorithm}
\author{Andrew D.~King}\email[]{aking@dwavesys.com}
\affiliation{D-Wave Systems Inc., Burnaby B.C.}
\author{William Bernoudy}
\affiliation{D-Wave Systems Inc., Burnaby B.C.}
\author{James King}
\affiliation{D-Wave Systems Inc., Burnaby B.C.}
\author{Andrew J.~Berkley}
\affiliation{D-Wave Systems Inc., Burnaby B.C.}
\author{Trevor Lanting}
\affiliation{D-Wave Systems Inc., Burnaby B.C.}
\date{\today}

\begin{abstract}
  The {\em coherent Ising machine} is an optical processor that uses coherent laser pulses, but does not employ coherent quantum dynamics in a computational role.  Core to its operation is the iterated simulation of all-to-all spin coupling via mean-field calculation in a classical FPGA coprocessor.  Although it has been described as ``operating at the quantum limit'' and a ``quantum artificial brain,'' interaction with the FPGA prevents the coherent Ising machine from exploiting quantum effects in its computations.  Thus the question naturally arises: Can the optical portion of the coherent Ising machine be replaced with classical mean-field arithmetic?  Here we answer this in the affirmative by showing that a straightforward noisy version of mean-field annealing closely matches CIM performance scaling, while running roughly 20 times faster in absolute terms.
\end{abstract}

\maketitle

\section{Introduction}

Advances in quantum information and the decline of Moore's Law have incited a flurry of research into quantum computing and other nontraditional computing schemes.  Among these schemes is the {\em coherent Ising machine} (CIM), prototypes of which exist at Stanford \cite{McMahon2016} and NTT \cite{Inagaki2016}.  The CIM is an optical machine used to solve problems in the Ising model using coherent laser pulses.  Despite the name, the CIM does not exploit multi-spin quantum dynamics in a computational role.  While it has been described as a ``quantum artificial brain'' \cite{NII2017} and ``operating at the quantum limit'' \cite{Yamamoto2017}, there is nothing ``quantum'' about it from a computational perspective.

The classical Ising problem solved by the CIM is a minimization of the energy function
\begin{equation}
H = \sum_{ij}J_{ij}\tilde s_i\tilde s_j + \sum_i h_i\tilde s_i,
  \end{equation}
where each $\tilde s_i = \pm 1$ is the sign of a continuous variable $s_i\in [-1,1]$ stored by the CIM in the phase of a laser pulse that circulates around a fiber optic loop.  Every round trip, the states of the pulses $s_i$ are measured and fed into an FPGA \footnote{A field-programmable gate array (FPGA) is a user-programmable integrated circuit used to quickly perform a specialized computational task.}%
. The FPGA calculates, for each $s_i$, the {\em mean field} imposed by the other pulses $\Phi_i = \sum_{j} J_{ij}s_j$.  The CIM then combines $\Phi_i$ with $s_i$ by generating a {\em mean-field pulse} via digital-to-analog conversion from the FPGA and optically combining it with $s_i$.

Since the pulses are repeatedly measured and are only connected indirectly through the FPGA, no entanglement between pulses is generated \cite{Yamamoto2017} and no useful quantum effects can survive beyond a single round trip.  In light of this, it is natural to ask whether we can replace the optical apparatus, which serves to store the spins and allow combination with a mean-field term, with an arithmetic operation in a conventional processor.

Here we compare the CIM with an analogous {\em noisy mean-field annealing} (NMFA) algorithm.  With its parameters fixed to a single set of values, NMFA closely matches the behavior of the CIM across a variety of instances studied in Refs.~\cite{Inagaki2016} and~\cite{Hamerly2018}.  NMFA attains similar success probabilities, but on a GPU runs roughly 20 times faster than the NTT CIM and 130 times faster than the Stanford CIM at the 100-spin scale, which is currently the maximum capacity of the Stanford CIM.

\section{Noisy mean-field annealing}

\begin{figure}
  \includegraphics[scale=.7]{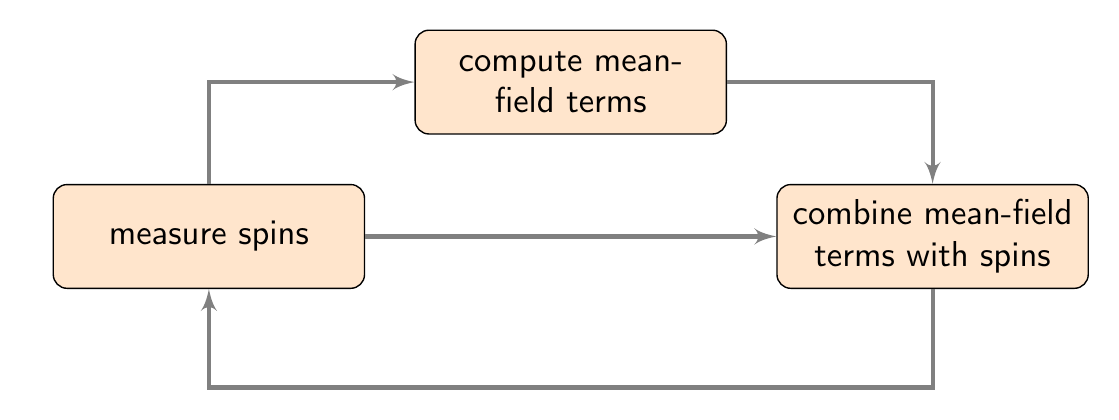}
  \caption{{\bf Algorithmic form of CIM and noisy mean-field annealing.}  Algorithmically, the CIM follows a simple loop in which spins are repeatedly measured, then combined with a mean-field term derived from that measurement.  We compare CIM performance with NMFA, an algorithm that follows the same loop using continuous real spin values in $[-1,1]$ instead of optical pulse phases.  While the CIM injects mean-field terms using optical pulses, NMFA injects mean-field terms using Boltzmann factors for a decreasing sequence of temperatures.}\label{fig:flowchart}
\end{figure}

\begin{figure}
\begin{algorithm}[H]
  \caption{Noisy mean-field annealing.  Generates a set of Ising spins $\tilde s_i$ given Ising problem $(h,J)$ and parameters $T$, $\sigma$, and $\alpha$.}\label{alg1}
  \label{NMFA}
  \begin{algorithmic}[1]
    \For{$i=1$ to $N$}
    \State $s_i := 0$
    \EndFor
    \For{$t=1$ to $t_f$}
    \For{$i=1$ to $N$}
    \State $\Phi_i := (h_i + \sum_j J_{ij}s_j) / \sqrt{h_i^2+ \sum_j J_{ij}^2   }+ \mathcal N(0,\sigma)$
    \State $\hat s_i := -\tanh{(\Phi_i/T_t)}$
    \EndFor
    \For{$i=1$ to $N$}
    \State $s_i := \alpha\hat s_i + (1-\alpha) s_i$
    \EndFor
    \EndFor
\For{$i=1$ to $N$}
    \State $\tilde s_i := s_i/|s_i|$
    \EndFor
    
   \end{algorithmic}
\end{algorithm}
   \includegraphics[scale=.8]{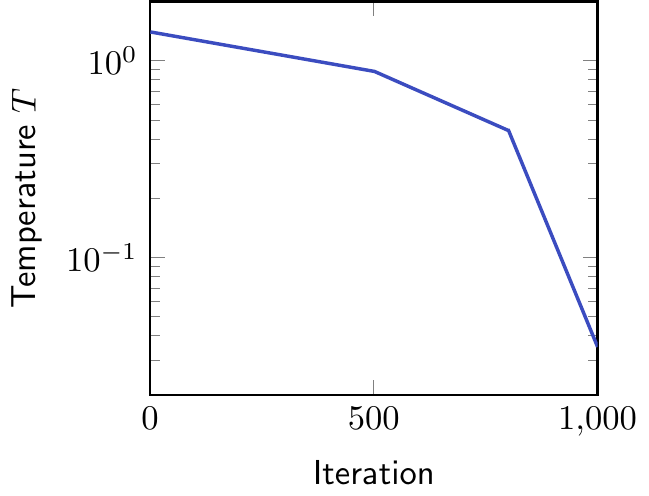}
\caption{{\bf Annealing schedule for NMFA.}  A three-segment piecewise exponential schedule provides good agreement between NMFA and CIM.}\label{fig:schedule}
\end{figure}

Broken down to its algorithmic form, the CIM implements a cycle of spin measurement, mean-field computation, and combination shown in Fig.~\ref{fig:flowchart}.  We implement the same cycle in a classical noisy mean-field annealing algorithm as follows.  Pseudocode is given in Algorithm \ref{alg1} and source code is provided as supplemental material.

\begin{figure}
  \includegraphics[scale=.8]{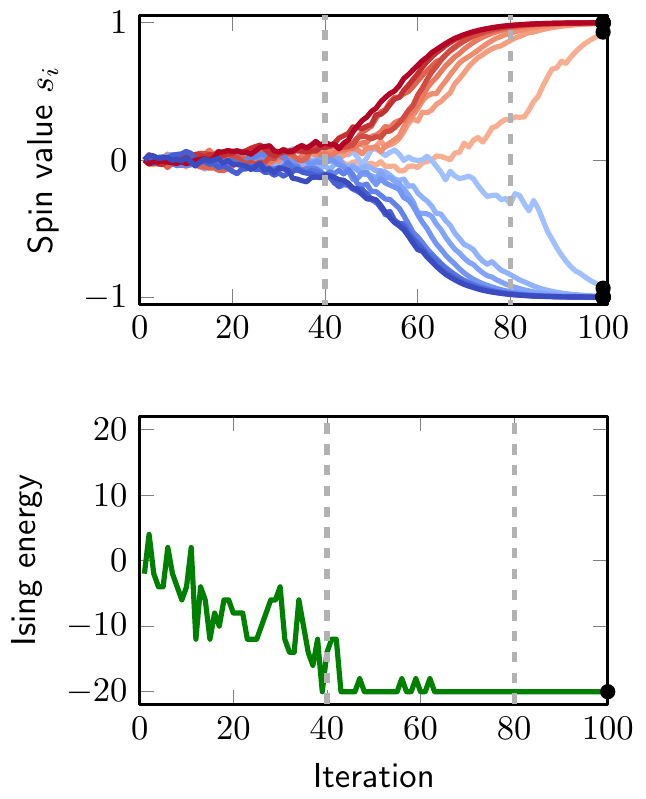}
  \caption{{\bf Example of NMFA run on a 16-spin M\"obius ladder.}  Evolution of spin values from 0 toward $\pm 1$ closely resembles behavior of the CIM on the same input (compare with Ref.~\cite{Hamerly2018} Fig.~1c).  Temperature schedule in Fig.~\ref{fig:schedule} is compressed to 100 iterations.}\label{fig:moebius}
\end{figure}

While the CIM uses the phase of an optical pulse to implement a spin $s_i$, NMFA stores spins $s_i$ as continuous values in the interval $[-1,1]$.  Computation of the mean-field terms is done in the same way for both solvers.  When it comes time to combine the mean-field term with the existing spin, the CIM does this by injecting a mean-field laser pulse into the existing spin pulse.  NMFA adds Gaussian noise with standard deviation $\sigma$ to the normalized mean-field term, then converts it to a spin value $\hat s_i$ using the Boltzmann expectation at a temperature $T$, i.e., $\hat s_i := -\tanh{(\Phi_i/T_t)}$.  NMFA then replaces spin $s_i$ with the convex combination $\alpha\hat s_i + (1-\alpha) s_i$.  The temperature decreases throughout the process.  This approach is essentially a noisy generalization of {\em mean-field annealing} \cite{Bilbro1989} with a feedback constant less than 1.

While NMFA is not intended to be a faithful microscopic model of the CIM, it nonetheless has all of the required ingredients: noisy analog spins, a mean-field feedback loop, and a means to smoothly evolve between noise-dominated and mean-field-dominated biases at the beginning and end of the computation, respectively.  Fig.~\ref{fig:moebius} shows the evolution of spin values $s_i$ (and the derived Ising energy) during a 100-iteration run on a 16-spin M\"obius ladder, showing very similar behavior to the CIM as shown in Ref.~\cite{Hamerly2018} Fig.~1c.

Note that in Line 6 of Algorithm \ref{alg1}, the mean-field term $\Phi_i$ is normalized by the root mean square of Hamiltonian terms acting on the spin, reflecting the expected magnitude of the mean field acting on that spin in a random state in the large-system limit.  In Line 7, $\hat s_i$ is computed as the Boltzmann expectation given $\Phi_i$ and $T_t$.

We used $\alpha=0.15$ and $\sigma=0.15$ for all experiments in this paper.  For all experiments except the 16-spin example in Fig.~\ref{fig:moebius} we used the annealing schedule shown in Fig.~\ref{fig:schedule}.  These parameters were tuned coarsely to show approximate agreement between NMFA and CIM.

\section{Results}

\begin{figure*}
  \begin{tikzpicture}
    \node[anchor=north west] at (0,0)
    {\includegraphics[scale=.8]{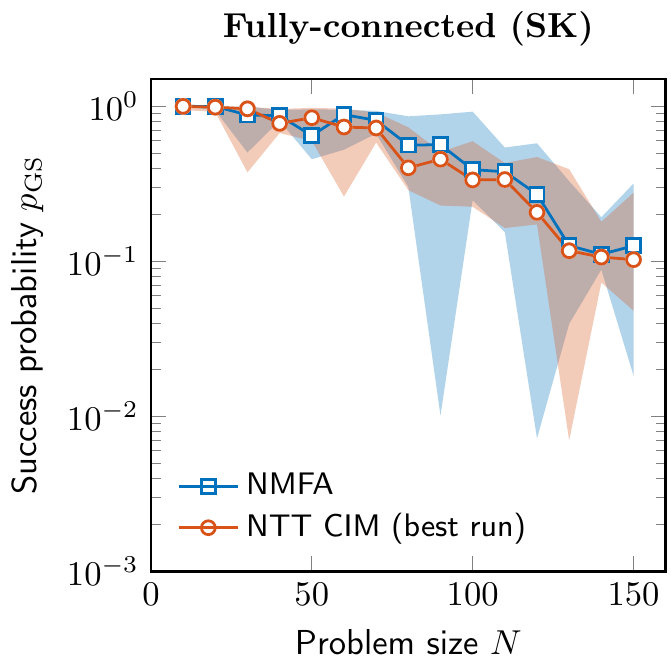}};
    \node[anchor=north west] at (6,0)
    {\includegraphics[scale=.8]{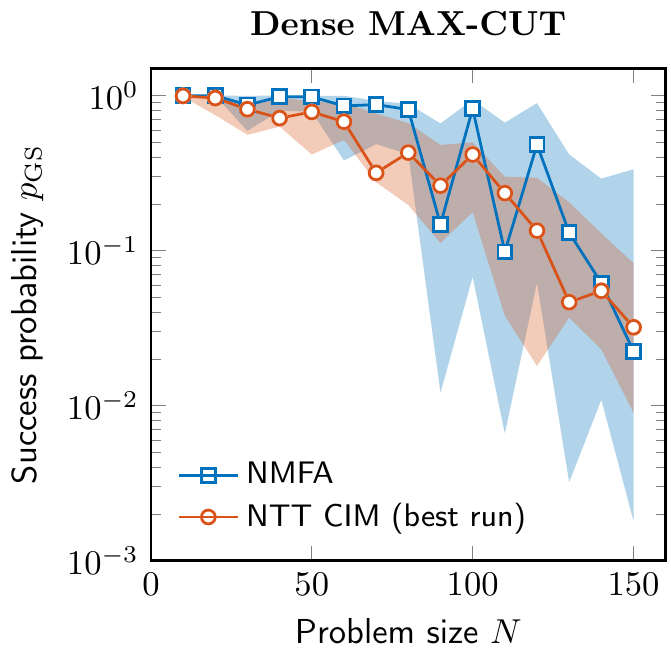}};
    \node[anchor=north west] at (12,0)
    {\includegraphics[scale=.8]{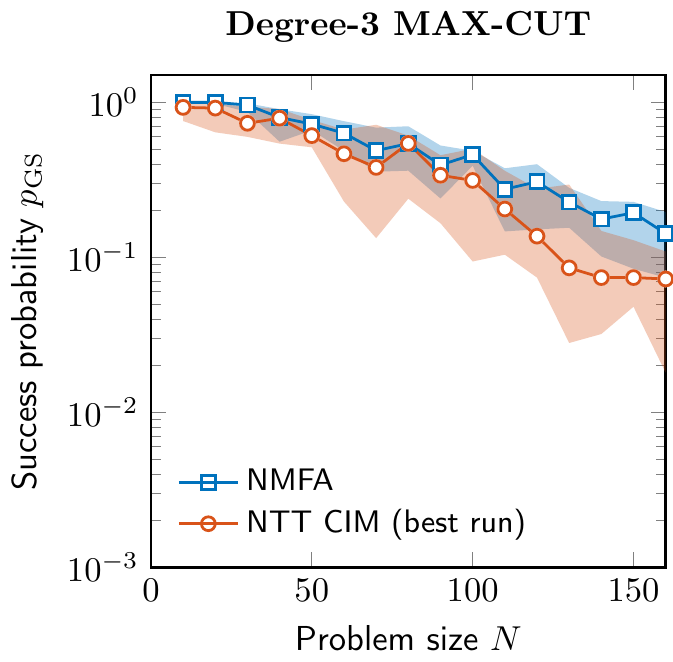}};
    \node[anchor=south west] at (0.3,-.8) {{\bfseries\sffamily\large a}};
    \node[anchor=south west] at (6.3,-.8) {{\bfseries\sffamily\large b}};
    \node[anchor=south west] at (12.3,-.8) {{\bfseries\sffamily\large c}};

    \node[anchor=north west] at (0,-5.5)
    {\includegraphics[scale=.8]{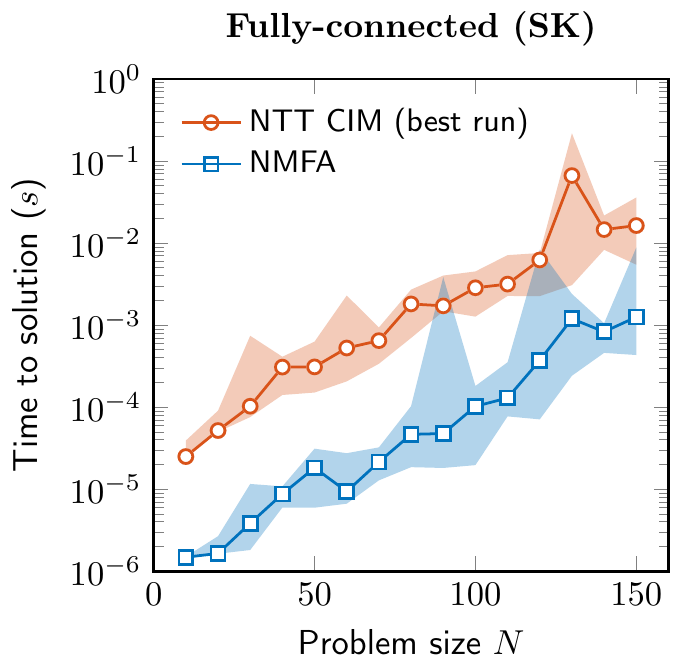}};
    \node[anchor=north west] at (6,-5.5)
    {\includegraphics[scale=.8]{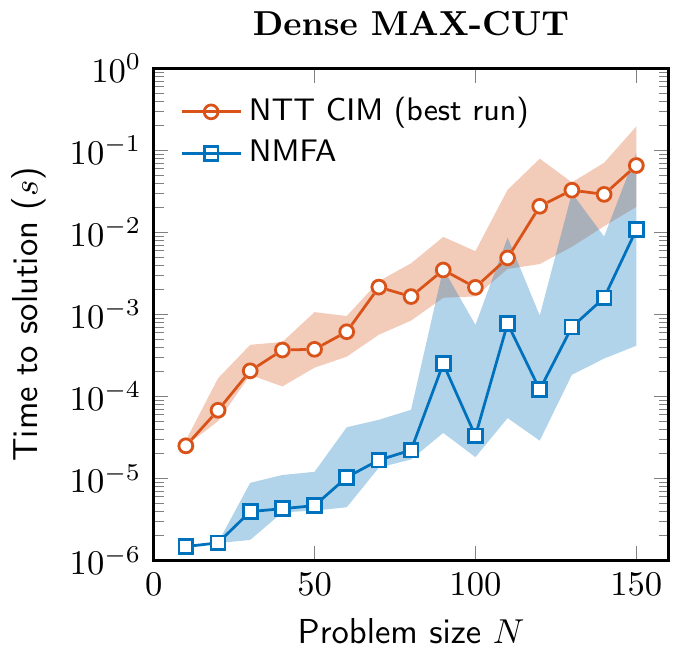}};
    \node[anchor=north west] at (12,-5.5)
    {\includegraphics[scale=.8]{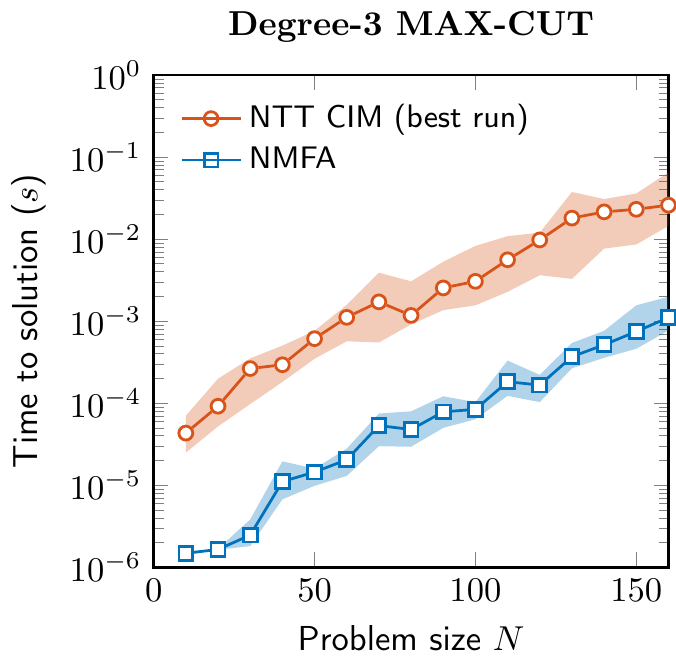}};
    \node[anchor=south west] at (0.3,-6.3) {{\bfseries\sffamily\large d}};
    \node[anchor=south west] at (6.3,-6.3) {{\bfseries\sffamily\large e}};
    \node[anchor=south west] at (12.3,-6.3) {{\bfseries\sffamily\large f}};
    
  \end{tikzpicture}
  \caption{{\bf Performance of NTT CIM and NMFA on problem sets studied in Ref.~\cite{Hamerly2018}.}  Marks and shaded regions represent medians and interquartile ranges. {\bf a--c}, success probabilities on ({\bf a}) fully-connected Sherrington-Kirkpatrick problems with $J_{ij} \in \{-1,1\}$, 10 instances per size;  ({\bf b}) dense MAX-CUT problems (edge probability $p=0.5$), 10 instances per size;  ({\bf c}) degree-3 MAX-CUT problems, 20 instances per size.  {\bf d--f}, time to solution.  NMFA time per sample is computed via wall-clock time across 10,000 samples; CIM time per sample is computed using $\SI{5}{ms}$ sample time for 2000 spins, divided by the number of copies of an instance that can be run in parallel.}\label{fig:results}
\end{figure*}

\begin{table*}\vspace{1cm}
\caption{{\bf Mean and best performance (out of 100 runs) on 2000-spin MAX-CUT instances detailed in Ref.~\cite{Inagaki2016}.}  Values given are for the maximization problem (MAX-CUT) rather than the equivalent Ising minimization problem.}\label{tab:table}
\begin{ruledtabular}
\begin{tabular}{cccc}
  Instance & $G22$ (random) & $G39$ (scale-free) & $K_{2000}$ (fully-connected)\\
  \hline
  NTT CIM \cite{Inagaki2016} &   mean 13248, best 13313  &  mean 2328, best 2361   & mean 32457, best 33191  \\
  NMFA & mean 13267,  best 13325   &   mean 2339,  best 2369   &  mean 32730, best 33186  \\
\end{tabular}
\end{ruledtabular}\vspace{1.35cm}
\end{table*}



\begin{figure*}
  \begin{tikzpicture}
    \node[anchor=north west] at (0,0)
    {\includegraphics[scale=.8]{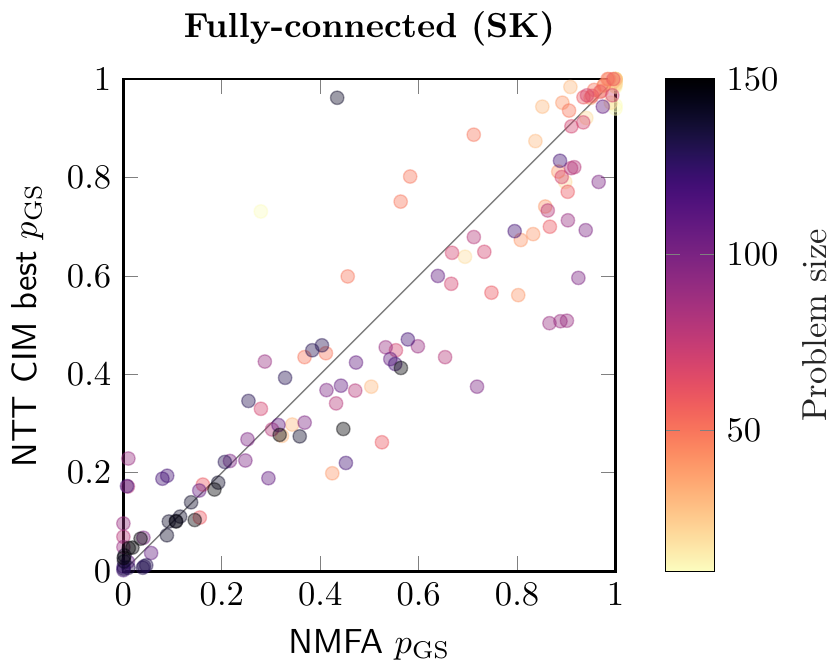}};
    \node[anchor=north west] at (9,0)
    {\includegraphics[scale=.8]{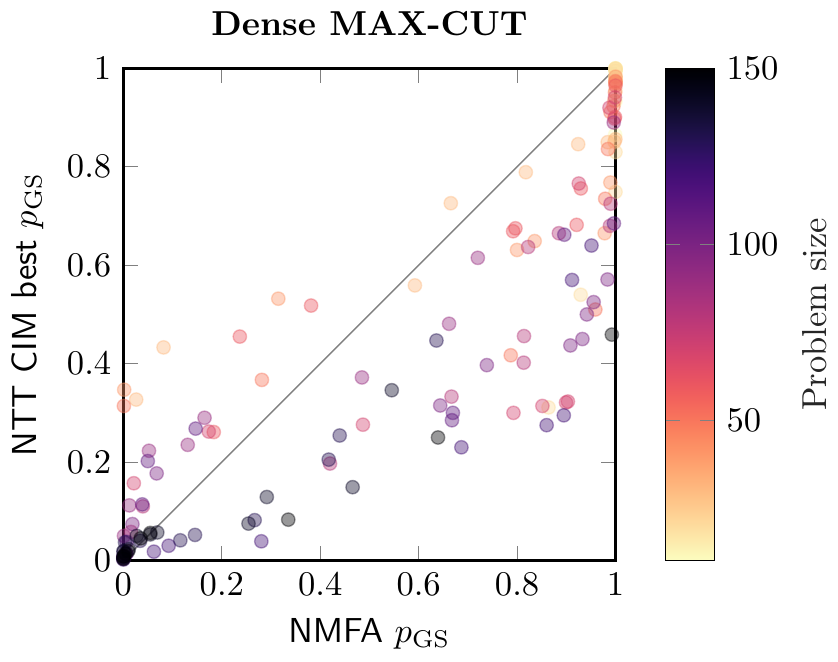}};
    \node[anchor=north west] at (5,-6)
    {\includegraphics[scale=.8]{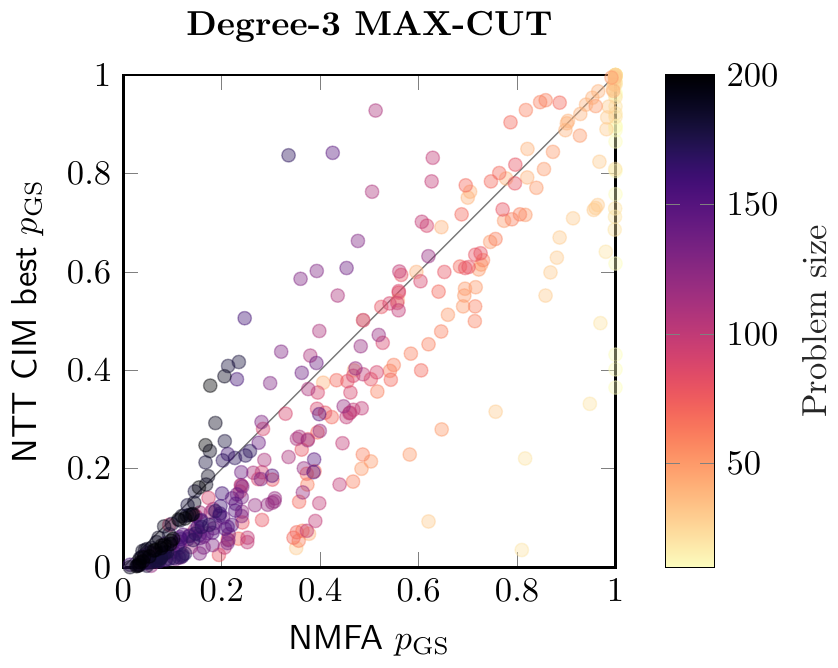}};f
    \node[anchor=south west] at (0.3,-.5) {{\bfseries\sffamily\large a}};
    \node[anchor=south west] at (9.3,-.5) {{\bfseries\sffamily\large b}};
    \node[anchor=south west] at (5.3,-6.5) {{\bfseries\sffamily\large c}};

  \end{tikzpicture}
  \caption{{\bf Instance-wise comparison of NTT CIM and NMFA.}  We compare success probabilities for each instance studied in Fig.~\ref{fig:results} for ({\bf a}) SK, ({\bf b}) dense MAX-CUT, and ({\bf c}) degree-3 MAX-CUT problems.}\label{fig:scatter}
\end{figure*}

Ref.~\cite{Hamerly2018} presents both Stanford CIM and NTT CIM results for several sets of random problems, which we reproduce in Fig.~\ref{fig:results} using NTT CIM and NMFA data.  In Fig.~\ref{fig:results}a, Sherrington-Kirkpatrick instances are studied, where $J_{ij}=1$ with probability $p=1/2$ and $J_{ij}=-1$ otherwise.  Fig.~\ref{fig:results}b shows results for dense MAX-CUT instances, where $J_{ij}=1$ with probability $p=1/2$ and $J_{ij}=0$ otherwise.  Fig.~\ref{fig:results}c shows data for degree-3 MAX-CUT instances, where each spin is coupled to three others (and all nonzero couplers have $J_{ij}=1$).  Fig.~\ref{fig:results}d--\ref{fig:results}f show performance in terms of time to solution using the common TTS metric used in Ref.~\cite{Hamerly2018} and elsewhere.  We show {\em best run} data for the CIM, where the best block of 1000 samples is chosen from 10,000 or more \cite{Hamerly2018}; for NMFA we show average performance for 10,000 samples.  In Fig.~\ref{fig:scatter} we give an instance-wise comparison.

Across these input classes, NMFA mirrors the success probabilities of CIM closely across the range of testbed parameters.  To get an idea of how long it takes each solver to draw a sample, we compare runtimes of CIM against NMFA for dense MAX-CUT instances with $N=100$, which is the maximum input size for the Stanford CIM.  The Stanford CIM takes $\SI{1600}{\micro\second}$ for a single run.  The NTT CIM, which has a maximum size of $N=2000$ and run time of $\SI{5000}{\micro\second}$, can run $20$ instances with $N=100$ in parallel, giving an effective run time of $\SI{250}{\micro\second}$.  NMFA, running on an NVIDIA GeForce GTX 1080 Ti GPU, takes $\SI{12.3}{\micro\second}$ per run, making it around 20 times faster than the NTT CIM and 130 times faster than the Stanford CIM, ignoring CIM overhead such as readout and postselection \cite{Hamerly2018}.

Table \ref{tab:table} shows mean and best MAX-CUT values over 100 runs for three 2000-spin instances studied in Ref.~\cite{Inagaki2016}, with NMFA using the same parameters as with the smaller instances.  On all three graphs, NMFA performs comparably to the CIM.

\section{All-to-FPGA-to-all connectivity}

Although the CIM is described as having all-to-all connectivity, there is no direct connection or coupling between any two spins.  Rather, all spins are measured and routed through an FPGA, where for each spin $s_i$ they are agglomerated as a single effective term $\Phi_i$, which is then output by the FPGA and routed back into the optical cavity.  This mean-field simulation of all-to-all connectivity could be achieved in a number of ways, including in a \mbox{D-Wave} processor by removing all inter-qubit connectivity and instead subjecting individual qubits to linear mean-field terms that evolve over a sequence of iterations.


\section{Conclusions}

The promise of any nontraditional computing method rests on it being able to do something faster, better, cheaper, or more easily than available methods.  In particular, it should be able to outperform a simple emulator running on a classical computer.  Our results show that on the instances studied so far, the coherent Ising machine falls short of this mark.

Similar criticism from outside researchers toward \mbox{D-Wave} quantum annealing processors  \cite{Smolin2013,Shin2014} ultimately led to advances in the field and more concrete validation of the quantum model in question \cite{Wang2013a,Lanting2014,Albash2015,Boixo2016,King2018}.  In the case of the coherent Ising machine, this outcome seems unlikely: while it does represent a novel use of optics, no macroscopic quantum model of computation has been proposed.  It is therefore unclear what, if any, potential utility the coherent Ising machine has as a future computing technology.

\section{Acknowledgments}

We thank Ryan Hamerly, Takahiro Inagaki, Ken-ichi Kawarabayashi, and Peter McMahon for useful discussions and for kindly providing experimental data.  We thank Richard Harris, Catherine McGeoch, Paul Bunyk, Jack Raymond, Isil Ozfidan, Emile Hoskinson, and Mohammad Amin for discussions on the manuscript.

\bibliography{cim}

\end{document}